\documentclass[conference]{IEEEtran}
\IEEEoverridecommandlockouts
\usepackage{cite}
\usepackage{amsmath,amssymb,amsfonts}
\usepackage{algorithmic}
\usepackage{graphicx}
\usepackage{textcomp}
\usepackage{xcolor}
\def\BibTeX{{\rm B\kern-.05em{\sc i\kern-.025em b}\kern-.08em
    T\kern-.1667em\lower.7ex\hbox{E}\kern-.125emX}}
    
\usepackage[lofdepth,lotdepth]{subfig}

\begin{document}

\title{GPU Sharing with Triples Mode
\thanks{DISTRIBUTION STATEMENT A. Approved for public release. Distribution is unlimited. This material is based upon work supported by the Under Secretary of Defense for Research and Engineering under Air Force Contract No. FA8702-15-D-0001. Any opinions, findings, conclusions or recommendations expressed in this material are those of the author(s) and do not necessarily reflect the views of the Under Secretary of Defense for Research and Engineering. © 2023 Massachusetts Institute of Technology. Delivered to the U.S. Government with Unlimited Rights, as defined in DFARS Part 252.227-7013 or 7014 (Feb 2014). Notwithstanding any copyright notice, U.S. Government rights in this work are defined by DFARS 252.227-7013 or DFARS 252.227-7014 as detailed above. Use of this work other than as specifically authorized by the U.S. Government may violate any copyrights that exist in this work.}
}

\author{\IEEEauthorblockN{Chansup Byun,  Albert Reuther, LaToya Anderson, William Arcand,  
Bill Bergeron, David Bestor,\\ Alexander Bonn,  Daniel Burrill, Vijay Gadepally, 
Michael Houle, Matthew Hubbell,\\ Hayden Jananthan, Michael Jones, 
Piotr Luszczek, Peter Michaleas, Lauren Milechin,\\ Guillermo Morales,
Julie Mullen, Andrew Prout, Antonio Rosa, Charles Yee, Jeremy Kepner 
}
\IEEEauthorblockA{
\textit{Massachusetts Institute of Technology}\\
}
}

\maketitle

\begin{abstract}
There is a tremendous amount of interest in AI/ML technologies due to the proliferation of generative AI applications such as ChatGPT.
This trend has significantly increased demand on GPUs, which are the workhorses for training AI models. 
Due to the high costs of GPUs and lacking supply, it has become of interest to optimize GPU usage in HPC centers. 
MIT Lincoln Laboratory Supercomputing Center (LLSC) has developed an easy-to-use GPU sharing feature supported by LLSC-developed tools including LLsub and LLMapReduce.
This approach overcomes some of the limitations with the existing methods for GPU sharing.
This allows users to apply GPU sharing whenever possible while they are developing their AI/ML models and/or doing parametric study on their AI models or executing other GPU applications.
Based on our initial experimental results with GPU sharing, GPU sharing with triples mode is easy to use and achieved significant improvement in GPU usage and throughput performance for certain types of AI applications. 
\end{abstract}


\section{Introduction}
There is a lot of interest in AI/ML technologies due to the proliferation of generative AI applications such as OpenAI ChatGPT~\cite{ChatGPT}.
Rapid growth in AI/ML related research, development, and commercial applications has increased demand for GPUs, which are the workhorse for training AI models.
Due to the high costs of GPUs and shortage of their supply, there is much greater interest in using GPUs more efficiently in HPC centers. 
Although some GPU applications require multiple GPUs to run because the application GPU memory requirement is too large to fit in a single GPU memory, not all GPU applications utilize all GPU memory nor require multiple GPUs.
For those GPU applications that cannot utilize a single GPU capacity in full, hardware vendors such as Nvidia and AMD have developed ways to share GPU resources among multiple users.
The NVIDIA Multi-Process Service (MPS)~\cite{NVIDIAmps} is a runtime architecture that is designed to transparently enable co-operative multi-process CUDA applications on a single GPU.
AMD's GPU virtualization approach~\cite{AMDmxgpu}, based on SR-IOV (Single Root, Input/Output Virtualization), allows up to 16 virtual partition per GPU.
Similarly, the NVIDIA Multi-Instance GPU (MIG) devices such as A100 cab split a single GPU into up to 7 multiple isolated GPU instances~\cite{NVIDIAmig}.

In recent years, several studies have been carried out about sharing GPU resources in order to increase GPU utilization.
A number of works related to the GPU virtualization are well documented in the reference~\cite{Ravi11}. 
Middleware-based solutions such as GPUShare~\cite{Goswami16} and SALUS~\cite{Yu20} focus on achieving fine-grained GPU sharing across multiple processes by enabling finer control of the time slice on the GPU and by enabling two GPU sharing primitives: fast job switching and memory sharing, respectively. 
Another approach is sharing unutilized GPU resources such as GPU registers and memory to launch more thread blocks which, in return, improves GPU performance~\cite{Jatala16}.
Furthermore, most commercial cloud service providers offer GPU sharing features using one or more of the following features: multi-instance, time-sharing, and multi-process service (MPS).~\cite{AmazonAWS, GoogleCloud, MSAzure}.

For HPC systems, Slurm supports NVIDIA MPS~\cite{NVIDIAmps} and MIG~\cite{NVIDIAmig} so that multiple jobs can share the same GPU on a cluster system~\cite{SLURMgres}.
However, NVIDIA MPS appears to have some limitations when used with Slurm, including that it is tricky to use when there are more than one GPUs on a compute node~\cite{NVIDIAmps}.
With NVIDIA MIG, Slurm can manage each GPU instance as an individual GPU resource to be managed by the scheduler.
Slurm also provides another mechanism to share a single GPU with multiple jobs, which is called GPU sharding~\cite{SLURMgres}. 
However, because there is no isolation between the processes running on the GPU, it can be a security issue when those jobs are owned by multiple users.
As described in the Slurm reference~\cite{SLURMgres}, it requires an advanced customization with Slurm configuration in order to share a GPU with multiple jobs.
This also requires additional Slurm options with job submission to request appropriate amount of GPU shares.

At MIT Lincoln Laboratory Supercomputing Center (LLSC), we have developed a different approach based on a unique LLSC environment.
LLSC systems now use a user-based whole-node scheduling policy  to address a number of issues that most shared HPC  systems may encounter. 
The user-based whole-node scheduling policy means that whole compute nodes are allocated to each user~\cite{Byun21, ByunSlurm23}.
In other words, once a user's job is dispatched to a compute node and there are unscheduled resources still available on that node, only other jobs from that same user can be scheduled on that node; jobs owned by other users cannot be scheduled on that node.
LLSC-developed tools such as LLsub, LLMapReduce~\cite{Byun16}, pMatlab/gridMatlab~\cite{Bliss07,Kepner09,Reuther04}, and pPython~\cite{Byun22, Byun23} support the whole-node scheduling policy with the triples mode~\cite{Byun21}.
These tools can easily control how many jobs are dispatched to each node and how each job is assigned to which GPU based on the GPU node specification. 
This feature can be exploited to share the given compute resources (either CPU or GPU) with multiple jobs.

This approach does not require any scheduler configuration customization. and it is easy to share CPU and/or GPU resources as needed.
For example, this allows users to apply GPU sharing whenever possible while they are developing their AI/ML models and/or doing parametric study on their AI models or executing other GPU applications.
Based on our initial experimental results, GPU sharing with the triples mode is easy to use and has demonstrated significant improvement in GPU usage and throughput performance for certain types of AI applications. 
In this paper, we describe how GPU sharing is implemented with the triples mode and present the performance improvements that GPU sharing has achieved with a couple of example ML applications.

\section{Triples Mode Extension}
As mentioned earlier, the sharing of computing resources is achieved by extending the triples mode~\cite{Byun21}, which is available with LLSC-developed tools. 
The triples mode (a.k.a., node-based job scheduling) enables the set of all the computing tasks from a user to be executed on a compute node to be scheduled as a single job with child tasks. 
This is accomplished by LLSC-developed tools by generating an execution script on the fly.
By doing so, the scheduling overhead is significantly reduced  and, in turn, it enables precise distribution of all of the tasks in a job using a triplet of integers that we call triples mode. 
Further, it enables the management of a large number of short running jobs which is not possible to handle with the typical job array~\cite{Byun21} because such situations often burden the scheduler to operate very slowly. 
The triples mode uses three integers to define how a set of parallel tasks is mapped onto and executed on a set of compute nodes, processes, and threads: total number of nodes ({\tt NNODE}), number of processes per node ({\tt NPPN}), and number of threads per process ({\tt NTPP}). 
The product of the first two numbers ({\tt NNODE} * {\tt NPPN}) equals the total number of processes (computing tasks) that are to be executed across all of the compute nodes.
The third argument define how many threads each of the processes is allowed to spawn per process using the OMP\_NUM\_THREADS environment variable.

For a normal job submission in which each process utilizes one thread, we recommend that users set the {\tt NPPN} number to be equal to the number of physical cores available on a compute node so that each process is scheduled to run on a single core.
For the GPU applications, the {\tt NPPN} number can be equal to the number of GPUs on a compute node so that each independent GPU application runs on a single GPU.
However, if a job using this normal job submission cannot utilize the compute resources sufficiently, we may consider an over-allocation of additional computing tasks in order to increase the resource utilization.
In other words, we would recommend that users increase the {\tt NPPN} number so that more than one task is utilizing each GPU simultaneously. 
In doing so, the triples mode code recognizes that multiple tasks are being assigned to each GPU on the compute node, and it round-robin assigns GPUs to the child tasks by assigning each individual process to a specific GPU by using the {\tt CUDA\_VISIBLE\_DEVICES} environment variable in the automatically generated execution script.  

To help users determine how much CPU load, GPU load, system memory, and GPU memory their jobs/tasks are using, users can monitor how compute resources are being utilized by using the {\tt LLload} command \cite{pearc-llload}.
If their jobs are under-utilizing compute resources, users may schedule additional computing tasks per node by increasing the {\tt NPPN} parameter, which can be determined by examining the memory usage and load of the CPU and/or GPU resources.  
For example, if the application is CPU based, the appropriate {\tt NPPN} number can be decided by looking at the CPU memory usage and CPU load.
For GPU applications, the GPU memory usage and GPU load are used to determine the appropriate {\tt NPPN} number. 
More details are included in the {\tt LLload} paper. 

\section{Experiments}
In order to demonstrate how GPU resource utilization can be increased by adopting GPU sharing with the triples mode, we have selected a couple of canonical machine learning (ML) models with their datasets, LeNet-4/MNIST~\cite{MNIST} and ResNet-18/ImageNet~\cite{ImageNet} using PyTorch \cite{pytorch}.

\subsection{MNIST}
\begin{figure*}[htbp]
   \centering
   \includegraphics[width=\textwidth]{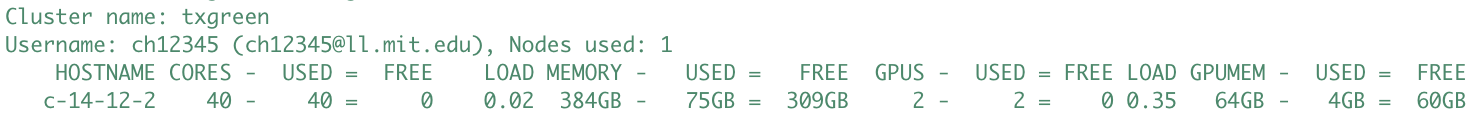}
%
%
  \caption{An \texttt{LLload} snapshot of resource usage when training the LeNet-4 ML model with the MNIST dataset using PyTorch.}
  \label{LLloadMNIST}
\end{figure*}

The MNIST dataset is a collection of images of handwritten numbers from 0 to 9, and the LeNet-4 model is the 4-layer convolutional neural network (CNN) model version that, along with LeNet-5, demonstrated a leap in performance of CNNs over traditional feed-forward neural networks on the numeral images of the MNIST dataset~\cite{LeCun1998}. 
This combination of model and dataset is a good example of an ML application that only modestly utilizes computational and memory capabilities on modern GPUs. 
In the training of the LeNet-4 model on MNIST using PyTorch with the default batch size of 64, only a small amount of memory is used for both CPU and GPU from the LLload output shown in Figure~\ref{LLloadMNIST}.
Also the CPU load is very low and the GPU load is modest, which indicates that the application does not utilize the many compute resources leaving a lot of the GPU capability idle.
It is obvious that a single training job cannot utilize the GPU resources in full.
This is a good opportunity to demonstrate to increase GPU utilization by sharing the GPU resources with multiple training jobs.

To demonstrate the effect of GPU sharing, we use the LLMapReduce command with the triples mode, a compute job of 24 identical MNIST training tasks is executed sequentially and concurrently on a single node which includes two NVIDIA Volta V100 GPUs.
In order to make the runtime to be reasonably long, we executed the training for 5 epochs.
The triples mode inputs used for this example are shown in Table~\ref{TriplesInput1}.
%
%
\begin{table}[htbp]
\caption{Triples Mode Inputs}
\begin{center}
\begin{tabular}{|c|c|c|c|}
\hline
\textbf{Concurrent}&\multicolumn{3}{|c|}{\textbf{Triples Mode Variables}} \\
\cline{2-4} 
\textbf{Jobs} & \textbf{\textit{NNODE}}$^{\mathrm{a}}$& \textbf{\textit{NPPN}}$^{\mathrm{b}}$& \textbf{\textit{NTPP}}$^{\mathrm{c}}$ \\
\hline
1& 1&  1&  40\\
2& 1&  2&  20\\
4& 1&  4&  10\\
6& 1&  6&  6\\
8& 1&  8&  5\\
12& 1&  12&  3\\
24& 1&  24&  1\\
\hline
\multicolumn{4}{l}{$^{\mathrm{a}}$Number of nodes, $^{\mathrm{b}}$Number of processes per node.}\\
\multicolumn{4}{l}{$^{\mathrm{c}}$Number of threads per process.}\\
\end{tabular}
\label{TriplesInput1}
\end{center}
\end{table}
The NTPP value is adjusted appropriately in order to not to overload the CPU resource as we increase the number of processes per node.
Although a single training job consumed about 4 GB of GPU memory as shown in Figure~\ref{LLloadMNIST}, we were able to run up to 24 training jobs concurrently on two Volta GPUs.
Each job is pinned to a specific GPU with the {\tt CUDA\_VISIBLE\_DEVICES} environment variable before executing the training job.

\begin{figure}[htbp]
   \centering
   \includegraphics[width=3.3in]{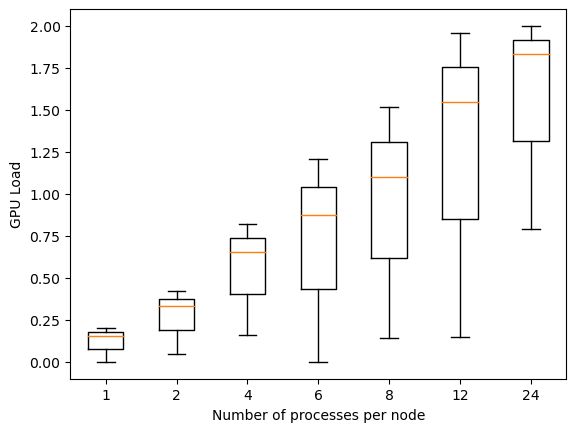}
  \caption{Observed GPU load distribution with respect to the number of concurrent training jobs.}
  \label{MNIST-gpu-load}
\end{figure}

GPU load and memory usage are monitored by running {\tt LLload} every 10 second as the experiment is running.
Figure~\ref{MNIST-gpu-load} shows how GPU load (minimum, average, and maximum load) changes as increasing the number of concurrent training jobs.
For each experiment with a given number of concurrent training jobs, GPU load varies while it is running over time.
However, the average GPU load clearly increases as the number of concurrent training jobs is increased.
This indicates that sharing the GPU resources with multiple jobs can be useful strategy to increase the GPU utilization.

\begin{figure}[htbp]
   \centering
   \includegraphics[width=3.3in]{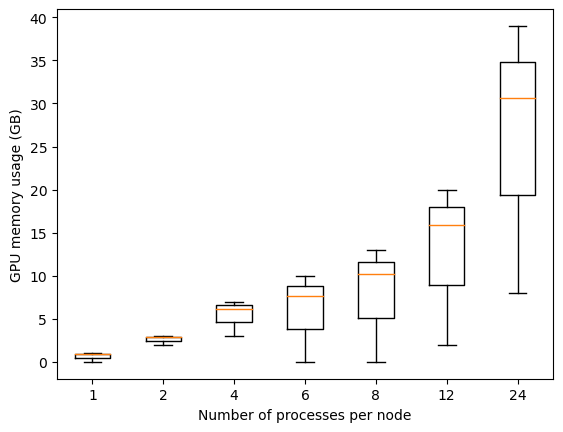}
  \caption{Observed GPU memory usage distribution with respect to the number of concurrent training jobs.}
  \label{MNIST-gpu-memory}
\end{figure}

Figures~\ref{MNIST-gpu-memory} shows how GPU memory usage changes as the number of concurrent training jobs is increased.
For each experiment with a given number of concurrent training jobs, GPU memory usage varies over time while it is monitored. 
However, the average GPU memory usage is clearly increasing as the number of concurrent training jobs is increased.
It is clear that, for up to 24 concurrent training jobs, the GPU memory requirement for the particular application did not exceed the maximum available GPU memory, though it came close.
However, as we will present in the next experiment, if GPU memory requirement exceeds the available GPU memory, the application will fail.
This means that tracking GPU memory usage is especially important when GPU is shared by multiple jobs.

\begin{figure}[htbp]
   \centering
   \includegraphics[width=3.3in]{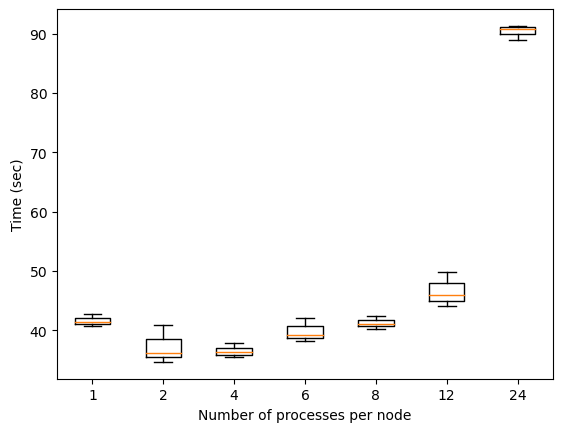}
  \caption{Individual training time variation with respect to the number of concurrent training jobs on a single node with two Volta 100 GPUs.}
  \label{MNIST-ind-time}
\end{figure}

Figure~\ref{MNIST-ind-time} shows how individual training time changes depending on how many jobs are running concurrently.
The timing data has been obtained from 24 training jobs per each concurrent job setting. 
For {\tt NPPN=1}, 24 training jobs are executed serially. (The second GPU remains idle in this instance.) 
For {\tt NPPN=2}, 2 training jobs are executed concurrently, one per each GPU and each GPU runs 12 training jobs sequentially.  
For {\tt NPPN=24}, 24 training jobs are executed concurrently, 12 jobs per each GPU.  
Increasing the number of concurrent processes (training jobs) usually increases the individual training time as well.  
However, there is a significant jump in individual training time when 24 training jobs are running concurrently.  
This is apparently too much of a load on GPU, and it slows down individual jobs significantly as a result.
With some likelihood, this may be caused by greater contention for accessing the DRAM memory of the GPUs, but further investigation is required. 

\begin{figure}[htbp]
   \centering
   \includegraphics[width=3.3in]{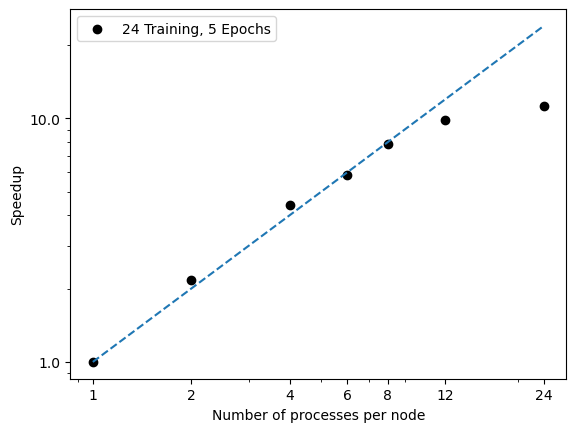}
  \caption{Speedup of the whole training job based on the job elapsed time with respect to the number of concurrent training jobs.}
  \label{MNIST-speedup}
\end{figure}

Overall throughput performance can be observed by calculating the speedup based on the elapsed time of each job with the given number of concurrent training jobs as shown in Figure~\ref{MNIST-speedup}.
Since this particular application needs only small amount of GPU memory, multiple training jobs can share the GPU resources to make full use of the GPU resources. As observed in Figure~\ref{MNIST-speedup}, linear speedup was observed up to 8 concurrent training jobs, 4 jobs per each GPU, and there is a slight efficiency drop with 12 concurrent training jobs. with 24 concurrent training jobs, a significant efficiency drop occurs. However, even with the significant efficiency drop at 24 concurrent training jobs, the overall training speedup of about 10 is still achieved, which is much faster compared to running the all of the training jobs sequentially.

We have also tried to run another experiment with 48 training jobs using the MNIST dataset. 
In this case, the experiment runs well up to 24 concurrent training jobs as observed in the previous experiment.
However, because running 48 concurrent jobs requires more GPU memory than the two Volta V100 GPU memory can provide, 21 of the 48 training jobs failed with CUDA out-of-memory errors.
Thus, it is important to look out for the GPU memory consumption when sharing GPUs with multiple jobs and make sure that the total memory usage by multiple jobs does not exceed the available GPU memory.

\subsection{ImageNet}

\begin{figure}[htbp]
\centering
\subfloat[][NPPN = 1]{
\includegraphics[width=2.5in]{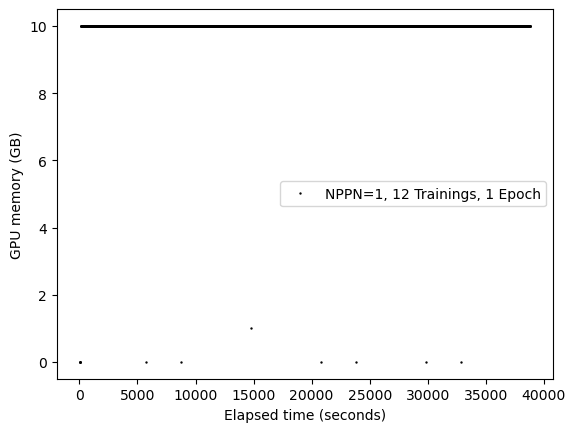}
\label{fig:subfig1}} 

\subfloat[][NPPN = 2]{
\includegraphics[width=2.5in]{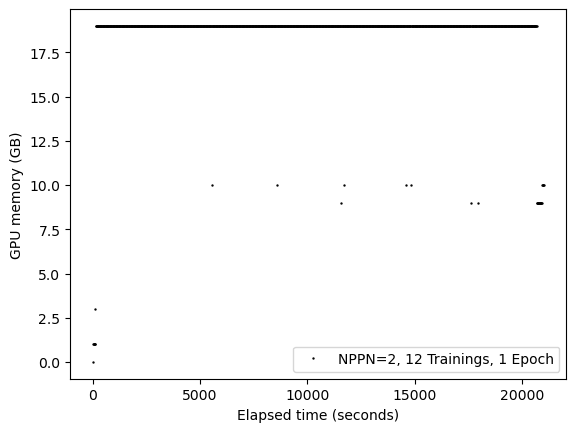}
\label{fig:subfig2}} 

\subfloat[][NPPN = 4]{
\includegraphics[width=2.5in]{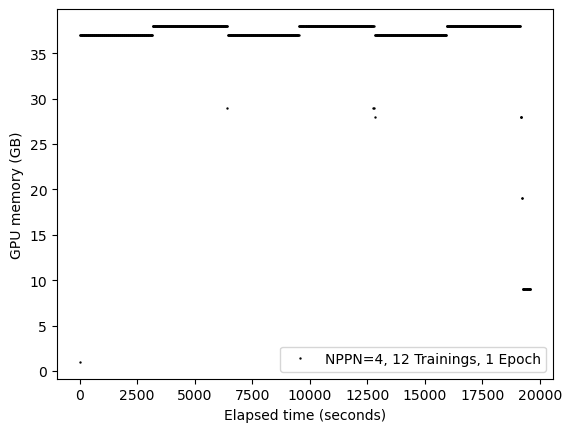}
\label{fig:subfig3}} 

\subfloat[][NPPN = 6]{
\includegraphics[width=2.5in]{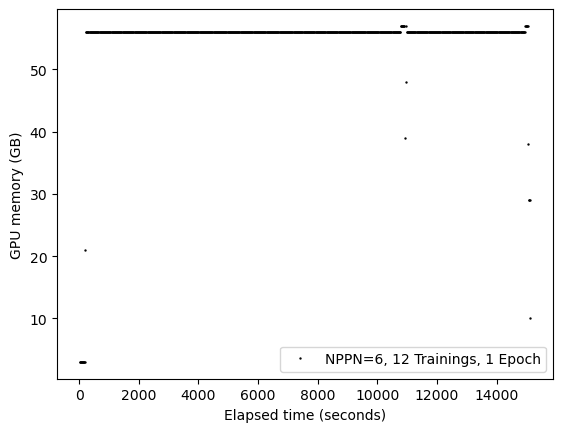}
\label{fig:subfig4}}
\caption{GPU memory usage history over time with various GPU sharing strategies, varying the number of concurrent processes per node ({\tt NPPN}).}
\label{fig:globfig}
\end{figure}
The ImageNet dataset contains 14,197,122 annotated images according to the WordNet hierarchy~\cite{ImageNet}. 
The ImageNet dataset has been used to train a wide variety of image detection and classification neural network models, and for this series of experiments, we used it to train ResNet-18 networks\cite{ResNet2016}. 
We have designed this experiment to run 12 individual training jobs serially and concurrently to demonstrate how GPU sharing behaves when each training job requires modest amount of GPU memory.
In order to monitor the GPU usage, the {\tt LLload -g} command is executed every 15 second and the snapshots are saved for analysis.
For this experiment, the training job is performed using the default batch size of 256 and default learning rate of 0.1 for one epoch.
LLMapReduce with the triples mode is again used to control different number of concurrent training jobs.
Based on the GPU memory usage, we have been able to perform up to 6 concurrent jobs per compute node, avoiding any CUDA out-of-memory errors.

Figure~\ref{fig:globfig} shows the GPU memory usage history over time with different number of concurrent processes per node (NPPN).
First, the peak GPU memory usage remained flat for each NPPN configuration, and it increases as NPPN is increased, since each training process maintains its own pool of GPU memory.  
With {\tt NPPN=6}, we observed that the peak GPU memory usage is close to the maximum GPU memory (64 GB) of two Volta GPUs.
Since there are total 12 training jobs to be performed, periodically we see sudden dips in GPU memory usage due to memory activities associated with job completions and subsequent job launches.
However, these sudden dips in GPU memory usage occur less frequently as NPPN is increased since more jobs are running concurrently and there are fewer job completions and subsequent job launches.

It is also worth noting that the total elapsed time to complete the entire training (total 12 jobs) task is significantly reduced from 38,848 seconds (~10.8 hours) for {\tt NPPN=1} to 15,136 seconds (~4.2 hours) for {\tt NPPN=6}.
This again demonstrates that GPU sharing increases GPU utilization and, in turn, increases overall throughput performance of the total training task almost 3x with {\tt NPPN=6} compared to {\tt NPPN=1}.

\begin{figure}[htbp]
\centering
\subfloat[][NPPN = 1]{
\includegraphics[width=2.5in]{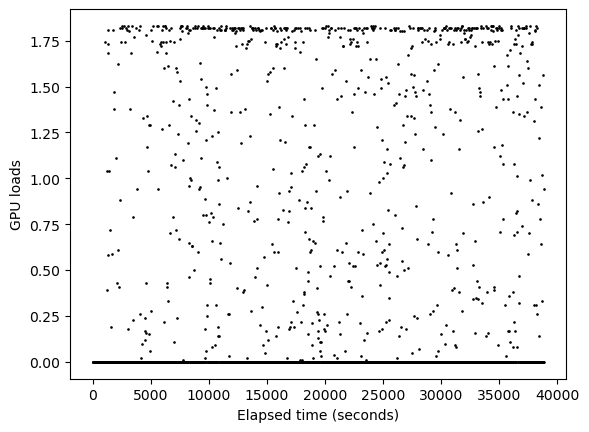}
\label{fig:in-load-subfig1}} 

\subfloat[][NPPN = 2]{
\includegraphics[width=2.5in]{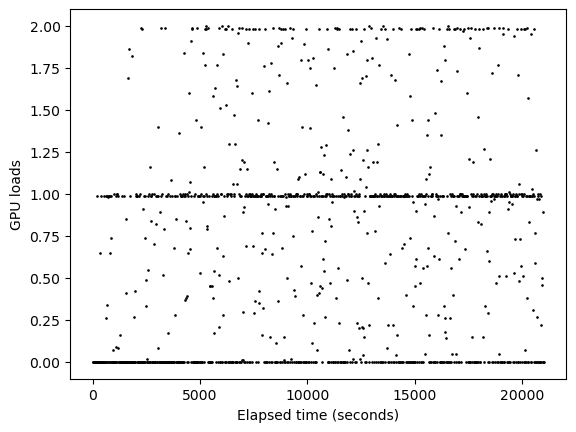}
\label{fig:in-load-subfig2}} 

\subfloat[][NPPN = 4]{
\includegraphics[width=2.5in]{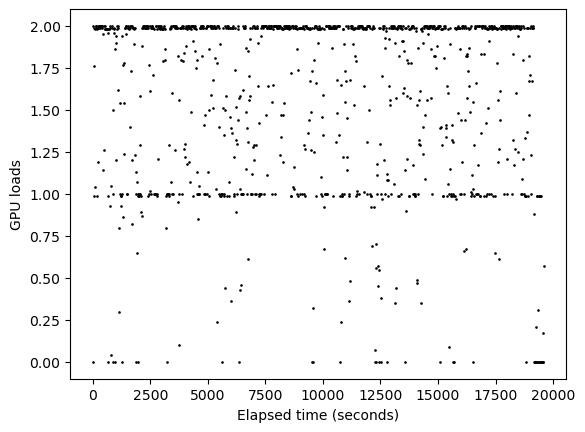}
\label{fig:in-load-subfig3}} 

\subfloat[][NPPN = 6]{
\includegraphics[width=2.5in]{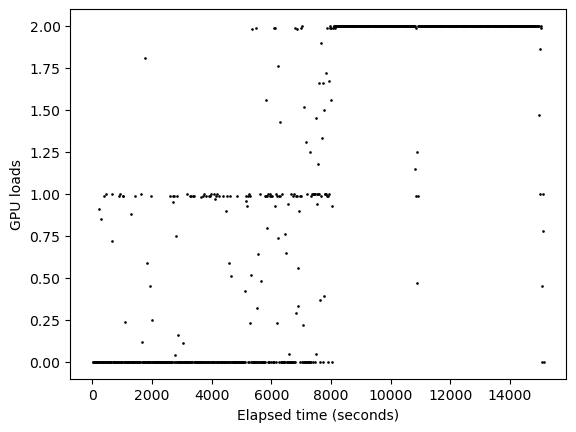}
\label{fig:in-load-subfig4}}
\caption{GPU load history over time with various GPU sharing strategies, varying the number of concurrent process per node ({\tt NPPN}).}
\label{fig:in-load-globfig}
\end{figure}
Figure~\ref{fig:in-load-globfig} shows how busy the GPUs are over time with varying number of concurrent processes per node.
Each point represents the GPU load snapshot at a point in time.
The GPU load is more scattered and sensitive than the GPU memory usage; the GPU load varies dramatically over time.
However, as NPPN is increased, the GPU load variation becomes less scattered, which indicates that the GPUs have more than enough work to be busy all the time. 
Further, this suggests that the GPU kernel queues have an adequate backlog of scheduled work that there are very few gaps between one computational kernel finishing execution on a GPU symmetric multiprocessor (SM) and another kernel beginning. 

There is a slight improvement in the total elapsed time when comparing the results between {\tt NPPN=2} and {\tt NPPN=4} in Figure~\ref{fig:globfig}.
This behavior can be explained by looking at the GPU load histories shown in Figures~\ref{fig:in-load-subfig2} and~\ref{fig:in-load-subfig3} -- their temporal GPU load history looked similar, but there is a higher concentration of GPU load along the GPU load = 1.0 line in~\ref{fig:in-load-subfig2} for {\tt NPPN=2}, while there is a higher concentration of GPU load along the GPU load = 2.0 line in~\ref{fig:in-load-subfig3} for{\tt NPPN=4}.
Please note that for {\tt NPPN=6} in Figure~\ref{fig:in-load-subfig4}, the first half part of the temporal GPU load is concentrated around 1.0 which is unexpected behavior; we are not sure what caused this behavior.
However, during the latter part of Figure~\ref{fig:in-load-subfig4}, the peak GPU load consistently hovered around 2.0, which helped result in a faster completion for the whole set of training tasks.

\begin{figure}[htbp]
   \centering
   \includegraphics[width=3.3in]{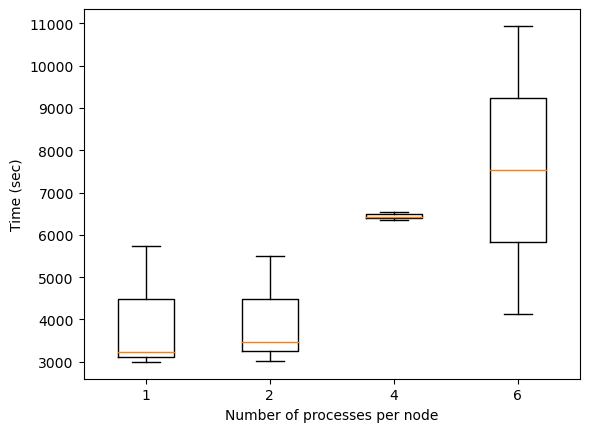}
  \caption{ResNet-18/ImageNet training time variation with respect to the number of concurrent training jobs on a single node with two Volta 100 GPUs.}
  \label{IN-ind-time}
\end{figure}

We also compared the individual training time of the 12 training jobs for each given number of concurrent processes ({\tt NPPN}) as shown in Figure~\ref{IN-ind-time}.
There is little change in the individual training time between {\tt NPPN=1} and {\tt NPPN=2}, and the variation of the individual training time looks similar between {\tt NPPN=1} and {\tt NPPN=2}.
This is expected since the particular PyTorch code is written for a single GPU resource and there are two GPUs on the node. 
However, there are significant increase in the individual training time with {\tt NPPN=4} and {\tt NPPN=6} because this involves GPU sharing between the assigned training jobs.
The variation of the individual training time becomes significantly larger with {\tt NPPN=6} because its GPU memory requirement is close to its physical GPU memory limit of 64 GB, and the GPUs are very busy as shown in Figure~\ref{fig:in-load-globfig}.
Again, there is likely main GPU memory contention between all of the training processes. 

\begin{figure}[htbp]
   \centering
   \includegraphics[width=3.3in]{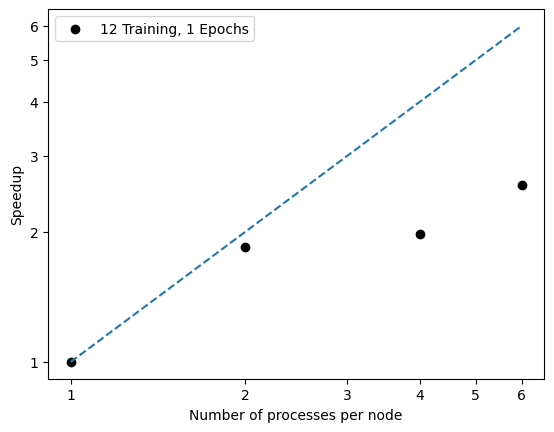}
  \caption{ResNet-18/ImageNet training by sharing the GPU resources with varying number of concurrent processes per node.}
  \label{IN-speedup}
\end{figure}

Finally, the speedup achieved by sharing the GPU resource with the ResNet-18/ImageNet training experiment is presented in Figure~\ref{IN-speedup}.
It shows almost linear speedup (1.85) from {\tt NPPN=1} to {\tt NPPN=2}, since the particular application is written for a single GPU, and the compute node has two Volta 100 GPUs.
But, it shows significant slowdown with GPU sharing with {\tt NPPN=4} and {\tt NPPN=6} where each GPU processes two and three concurrent training jobs respectively.
However, overall throughput performance has been improved substantially from {\tt NPPN=1} to {\tt NPPN=6} by 2.56 times based on the elapsed time to complete the whole training jobs.
It is obvious but still important to understand the application in order to get the best use of the GPU resources.
All of these experiments show that GPU sharing can achieve better throughput performance and reduce the total elapsed time for the given training jobs, even for applications in which each task shows modest GPU memory usage.

\section{Summary}
LLSC has developed an easy-to-use resource sharing tool by extending the triples mode.
This study is focused on GPU sharing which enables the launching of multiple ML applications on each GPU, which in return increases GPU utilization and overall throughput performance.
By launching multiple processes, it is possible to minimize any idle time of GPU resources.

The LLSC approach does not requires any sophisticated middleware implementation nor the customization of the scheduler configuration.
Thus, it is easy to deploy on the production environment.
In addition, this approach allows users to apply GPU overloading whenever possible while they are developing their AI/ML models, doing parametric study on their AI models, or executing other GPU applications.
Based on our initial results with GPU sharing experiments, GPU sharing with the triples mode is beneficial for certain types of AI applications and make more efficient use of GPUs in HPC environment and, in turn, increases overall throughput significantly.

\section*{Acknowledgment}
The authors express their gratitude to Bob Bond, Alan Edelman, Jeffrey Gottschalk, Charles Leiserson, Kristen Malvey, Heidi Perry, Stephen Rejto, Mark Sherman and Marc Zissman for their support of this work.

\bibliographystyle{IEEEtran} 
\bibliography{HPEC_2024_GPU_sharing}

\end{document}